\documentclass[nonacm,sigconf]{acmart}
\usepackage{graphicx}

\acmISBN{978-1-4503-XXXX-X/18/06}

\begin{document}

\title{Is Stateful Fuzzing Really Challenging?}

\author{Cristian Daniele}
\email{cristian.daniele@ru.nl}
\affiliation{%
  \institution{Radboud University}
  \city{Nijmegen}
  \state{}
  \country{Netherlands}
}









\begin{abstract}
  Fuzzing has been proven extremely effective in finding vulnerabilities in software.
 When it comes to fuzz stateless systems, analysts have no doubts about the choice to make. In fact, among the plethora of stateless fuzzers devised in the last 20 years, AFL (with its descendants AFL++ and LibAFL) stood up for its effectiveness, speed and ability to find bugs.
 On the other hand, when dealing with stateful systems, it is not clear what is the best tool to use. In fact, the research community struggles to devise (and benchmark) effective and generic stateful fuzzers. In this short paper, we discuss the reasons that make stateful fuzzers difficult to devise and benchmark. 
  \end{abstract}



\keywords{Stateless fuzzing, Stateful fuzzing, Software testing}


\maketitle

\section{Introduction}\label{sec:intro}

Fuzzing consists of sending many malformed (or slightly malformed) messages to a program in order to trigger crashes. Despite the approach being straightforward, clever engineering stunts (like code instrumentation) made the technique very effective and scalable. The pioneer of the approach is AFL~\cite{afl} (and its successors AFL++ and the more recent LibAFL~\cite{fioraldi2022libafl}), which, since 2013, has found plenty of bugs daily.
The problem is that the above-mentioned tools are devised to tackle stateless systems (as explained in Section~\ref{sec:differences}), i.e. systems that do not have any notion of sessions and do not need to implement any state model. 
Unfortunately, the vast majority of protocols --- being stateful --- need state models to work. 
The statefulness of these systems makes \textit{all} the stateless fuzzers almost totally ineffective (as explained in Section~\ref{sec:limits-of-stateless-fuzzers}). For this reason, the community developed fuzzers able to bypass or deal with the statefulness of the systems (as explained in Section~\ref{sec:stateful-fuzzers}).

\section{Stateless and Stateful Fuzzing}\label{sec:differences}

Fuzzers (both stateless and stateful) send millions of messages to the System Under Test (SUT) to trigger crashes or weird system behaviors~\cite{zhu2022fuzzing}. To generate these messages cleverly, mutational-based fuzzers (often called smart evolutionary or grey-box), instrument the code to get feedback from the SUT about the quality of the messages sent. Basically, the fuzzer rewards messages that trigger new coverage in the SUT code. 

The main difference between stateless and stateful fuzzers is that while the formers only need to mutate the single messages to (theoretically) get 100\% coverage, the latter also need mutations over the order of the messages (often called trace) to (theoretically) get 100\% coverage. For this reason, stateless fuzzers usually achieve poor results fuzzing stateful systems.

\section{Limits of stateless fuzzers}\label{sec:limits-of-stateless-fuzzers}
Despite being possible to use stateless fuzzers over stateful systems, it is usually not a good choice.
In fact, stateless fuzzers lack some abilities required to efficiently fuzz stateful systems. Namely, they cannot:
\begin{enumerate}
  \item send messages over the network. It is easy to implement, but requires modifing the fuzzer's code or using specific libraries~\cite{andarzian2023green};
  \item avoid the SUT from restarting after every message. AFL and all the AFL-based fuzzers restart the SUT after every message. This behaviour precludes the possibility of fuzzing deep states since, after the restart, the SUT starts again from the initial state;
  \item mutate the order of the messages sent to the SUT. Stateless fuzzers do not provide mutation functions on the trace level, making impossible mutations over the traces;
  \item focus on the most interesting state (or any state, altogether), as they do not have any notion of state models;
  \item observe the response of the server to steer the generation of the messages or infer the state model of the SUT.
\end{enumerate}

The above-mentioned flaws highlight the need to devise fuzzers able to deal with stateful systems.

\section{Approaches used by stateful fuzzers} \label{sec:stateful-fuzzers}

Different approaches have been developed to fuzz stateful systems. While some of them try to bypass the stateful nature of the system; others try to deal with it.

\subsection{Bypassing the statefulness of the systems}
The research community came up with different approaches to using stateless fuzzers over stateful systems.
\subsubsection{Bringing the SUT to a certain state}
The simplest approach to bypass the statefulness of the systems is perhaps bringing the SUT to a certain state (by using a prefix) and then using a stateless fuzzer. For example, the \textit{custom mutators} implemented by AFL++ easily allow prepending fixed prefixes to the messages to mutate. 

In order to use a more systematic approach, it is possible to describe the prefixes through a grammar and use a grammar-based fuzzer. For a grammar-based fuzzer (\cite{gascon2015pulsar}, ~\cite{doupe2012enemy}) the statefulness of a system is not an issue as the grammar provided can describe both the structure of the messages (needed also to fuzz stateless systems) and the structure of the traces~\cite{daniele2023fuzzers}. As already mentioned, the grammar of the state model is often \textit{only} used to lead the SUT to a specific state. In other words, the fuzzers do not use the grammar for the mutations but to reach and tackle interesting\footnote{In this scenario, the analyst decides whether a state is interesting or not.} states. The reason behind this choice is straightforward: mutations over the content and the order of the messages will likely lead to chaotic trace generations. 

\subsubsection{Adding artificial loops}
Another approach to bypass the statefulness of the systems consists of adding an artificial loop (in the SUT) that wraps the handling of the commands. This loop avoids the restarting of the SUT after every message and --- very artificially --- allows mutations over the traces.
\textit{AFL persistent mode} (originally devised exclusively stateless systems) allows such a behaviour. It allows annotating the SUT code with a \textit{AFL\_LOOP(n)} instruction to cycle a specific portion of code $n$ times~\cite{anonymous2024afl*:}. A drawback of the approach is the possibility of ending up in an unprofitable state (such as a state that merely handles errors). To avoid this problem, it is possible to include in the fuzzing dictionary (often called seed file) a message that brings back to the initial state.\footnote{The SUT needs to implement such a command; otherwise the analyst needs to implement it.}

\subsubsection{Using fuzzing targets}
To bypass the statefulness of the systems, it is also possible to use \textit{fuzzing targets}. Stateless fuzzers like LibFuzzer allow one to tackle portions of code by providing specific entry points to specific functions. Despite this approach theoretically working in fuzzing stateful systems, it can overlook all the bugs triggered by some sort of relation between the different messages in the same trace. For example, assume that the trace $<m_{1},m_{2},m_{3}>$ triggers the bug. Thanks to the fuzzing targets, it is possible to fuzz specifically the piece of code parsing the message $m_{2}$. Unfortunately, this will overlook all the issues caused by weird relations between the different messages in the trace. Moreover, this approach makes it challenging to trigger the bug from the outside as the analyst has no clue about the structure of the whole trace. In fact, analysts would only know the single message that triggers a bug in that specific function.

\subsubsection{Cram multiple messages into one}
Another strategy to bypass the statefulness of the systems by using an AFL-based fuzzer is to cram several messages (separated by a delimiter) into one. The message would then be something like $M={m_{1}\backslash nm_{2} \backslash nm_{3}}$, assuming the delimiter being the newline character ($\backslash n$). Doing this allows fuzzing deeper states but does not give any control over the mutations of the single messages. In fact, mutating the single messages would require ad hoc mutation functions able to chop the messages up. To the best of our knowledge, no fuzzer uses this approach. 
\bigskip

It is worthwhile to note that all the above-mentioned approaches only solve the problem (2) described in Section~\ref{sec:limits-of-stateless-fuzzers} but still do not allow focusing on the more promising states or leveraging server responses.

\subsection{Dealing with the statefulness of the systems}
  
Other approaches (\cite{pham2020aflnet},~\cite{ba2022stateful},~\cite{natella2022stateafl}), instead of trying to bypass the statefulness of the systems, try to deal with it by sending \textit{entire traces} (and not single messages) to the SUT. 

This allows the fuzzers to make mutations over single messages and also over their order. More advanced fuzzers also have modules to infer and take into account the state model of the SUT. 
Sending entire traces is slow. For this reason, some stateful fuzzers (like nyx-net~\cite{schumilo2022nyx}) introduced a technique called \textit{snapshotting} that allows to \textit{snapshot} the program in a certain state to be able to reach the same state more quickly in the future. 
Despite the technique aiming to make the fuzzing process faster, it often introduces such a huge overhead that makes re-sending the entire sequence of messages more convenient.

Unfortunately, the majority of these fuzzers are tailored to specific systems and require some tuning to work nicely on others. Typical modifications required are the writing of modules to parse protocols requests and responses~\cite{pham2020aflnet}, and SUT modifications to highlight state variables~\cite{ba2022stateful}, getting rid of forks~\cite{maugeri2023evaluating}, sockets~\cite{andarzian2023green}, and any cryptography whatsoever.

\section{Benchmarking stateful fuzzers}
Another issue when coming to stateful fuzzing regards their benchmarking. In fact, while for stateless fuzzers it is reasonable to benchmark them according to the code coverage they achieved, things get more complicated with the stateful ones.
For stateful fuzzers, also the \textit{state coverage} plays a crucial role since high coverage in the code does not imply high coverage in the state model. Unfortunately, monitoring the state coverage is challenging and requires the actual state model of the SUT to track which (and how many times) states have been fuzzed. 

To the best of our knowledge, ProFuzzBench~\cite{natella2021profuzzbench} is the only benchmark framework for stateful protocols. Despite being widely used to compare stateful fuzzers, it only compares fuzzers regarding \textit{code coverage}. Useful extensions to the framework might be modules to monitor the state coverage and more fine-grained information about the number of messages sent and time in relation to the coverage. 
Monitoring the time needed to achieve a certain coverage only gives information about the speed of the fuzzer. On the other hand, coverage in relation to messages sent gives information about the ability of the fuzzer to craft interesting messages. Minimising the number of messages sent implies a deeper understanding of the structure of the SUT.

Another challenge when benchmarking stateful fuzzers regards the possible inconsistency of some stateful fuzzers. In fact, some fuzzers might be good at exploring the state model while others might excel in fuzzing single states, so the benchmarking can be highly biased by the nature of the system tested (i.e. SUT implementing simple or complex state models).

\section{Conclusions}\label{sec:conclusions}

Despite the community being actively involved in devising stateful fuzzers, most of the approaches are not scalable but tailored to particular SUTs. The challenges presented in the paper show that \textit{the} stateful fuzzer is still far away. The lack of mature and generic stateful fuzzers is partially justified by stateful fuzzing being a much younger field than stateless fuzzing. 

Eventually, this paper sheds light on the challenges and techniques of stateful fuzzing, giving room for further research and future directions.

\begin{acks}
  This research is funded by NWO as part of the INTERSECT project
  (NWA.1160.18.301).
\end{acks}
\bibliographystyle{ACM-Reference-Format}
\bibliography{bib}

\end{document}